\documentstyle[12pt, epsfig]{article}
\newcommand{\bea}   {\begin{eqnarray}}
\newcommand{\eea}   {\end{eqnarray}}
\begin{document}
\renewcommand{\thefootnote}{\fnsymbol{footnote}}

\thispagestyle{empty}
\title{On Alphabetic Presentations of Clifford Algebras and Their Possible Applications}
\author{Francesco Toppan\thanks{{\em e-mail: toppan@cbpf.br}}
~and Piet W. Verbeek\thanks{{\em e-mail: P.W.Verbeek@tudelft.nl}}
~\\~ \\
{\it $~^{\ast}$CBPF, Rua Dr.}
{\it Xavier Sigaud 150,}
 \\ {\it cep 22290-180, Rio de Janeiro (RJ), Brazil.}\\
 {\it $~^\dagger$Delft University of Technology, Lorentzweg 1,}\\{\it 2628 CJ Delft, The Netherlands.}}
\maketitle
\begin{abstract}
In this paper we address the problem of constructing a class of representations of Clifford algebras that can be named ``alphabetic (re)presentations''. The Clifford algebras generators are expressed as $m$-letter words written with a $3$-character or a $4$-character alphabet.
We formulate the problem of the alphabetic presentations, deriving the main properties and some general results. At the end we briefly discuss the motivations of this work and outline some possible applications.
\par

\end{abstract}
\vfill
\rightline{CBPF-NF-001/09}

\newpage
\section{Introduction}
The irreducible representations of Clifford algebras have been classified in \cite{ABS}. Convenient reformulations of this result can be found, e.g. in \cite{oku} and \cite{por}, where some topics, like the connection with division algebras, are also discussed.\par
The $Cl(p,q)$ Clifford algebra over the real is the enveloping algebra generated by the $\gamma_i$ real matrices ($i=1,\ldots, p+q$) and quotiented by the relation
\bea\label{gammabasis}
\gamma_i\gamma_j+\gamma_j\gamma_i&=&2\eta_{ij}{\bf 1},
\eea
where $\eta_{ij}$ is a diagonal matrix with $p$ positive entries $+1$ and $q$ negative entries $-1$.
In the following a basis of $p+q$ gamma matrices $\gamma_i$ satisfying (\ref{gammabasis}) will be called a gamma basis.\par
The real irreducible representations are, up to similarity transformations, unique for $p-q\neq 1,5~mod ~8$
while, for $p-q=1,5~mod~8$, there are two inequivalent irreducible representations which can be recovered by flipping the sign
($\gamma_i\mapsto -\gamma_i$) of all gamma basis generators. The size $n$ of an $n\times n$ real matrix irreducible representation is specified in terms of $p$ and $q$.\par
Both in \cite{oku} and \cite{crt}, given gamma basis representatives of a $Cl(p,q)$ real irreducible representation
 were explicitly constructed (for any $p,q$ pair), up to an overall sign flipping, in terms of tensor products of four basic $2\times 2$ real matrices. In \cite{crt} the four matrices were named $\sigma_1$, $\sigma_2$, $\sigma_A$, $\bf 1$ and defined as follows
\bea\label{fundmat}
\sigma_1=
\left(\begin{array}{cc}
	0&1\\
	1&0
\end{array}\right), &&\sigma_2=\left(\begin{array}{cc}
	1&0\\
	0&-1
\end{array}\right),\nonumber\\
{\bf 1}_2=
\left(\begin{array}{cc}
	1&0\\
	0&1
\end{array}\right), &&{\sigma}_A=\left(\begin{array}{cc}
	0&1\\
	-1&0
\end{array}\right).
\eea
Without loss of generality, e.g., the three irreducible gamma generators of, let's say, $Cl(3,0)$, can be explicitly given by
\bea
&\gamma_1={\bf 1}\otimes \sigma_1, ~~\gamma_2={\bf 1}_2\otimes \sigma_2, ~~\gamma_3=\sigma_A\otimes \sigma_A&
\eea
(any different presentation for the $Cl(3,0)$ gamma basis is equivalent by similarity). \par
Extending this result, the $p+q$ generators of a given real irreducible $Cl(p,q)$ Clifford algebra can be expressed as strings of tensor products of the $4$ matrices above, taken $m$ times (if $n$ is the size of the irreducible representation, therefore $n=2^m$; in the previous $p=3$, $q=0$ example, $n=4$ and $m=2$).\par
In the above type of gamma basis presentations, a few points should be noticed. At first the introduction of the
tensor product symbol ``$\otimes$'' is redundant. Once we understood that we are dealing with tensor products,
we do not need to write it explicitly. For the same reason, the four matrices given in (\ref{fundmat}) can be
expressed with $4$ characters of some given alphabet. For our purposes here we choose the four characters being given by $I,X,Z,A$; we associate them to the above gamma matrices according to
\bea\label{translation}
&{\bf 1}_2 \equiv I,~~{\sigma_1}\equiv X,~~ \sigma_2\equiv Z,~~ \sigma_A\equiv A,&
\eea
(``$A$'' stands for antisymmetric, since $\sigma_A$ is the only antisymmetric matrix in the above set).
\par
In the example above the three gamma matrices $\gamma_i$ can be more compactly expressed through the positions
\bea\label{alpha30}
&\gamma_1\equiv IX, ~~\gamma_2\equiv IZ, ~~\gamma_3\equiv AA.&
\eea
With the above identifications, for any $(p,q)$ pair (with the exception of the trivial $p=1$, $q=0$ case) and
up to an overall sign factor, we can always write down the $p+q$ generators of a gamma basis as $m$-letter words (the value $m$ is common to all words of the basis), written with the four $I,X,Z,A$ characters. For obvious reasons
we call this type of gamma matrix presentations ``alphabetic presentations'' or ``alphabetic representations'', according to the context. \par
Not all representations are alphabetic according to the previous definition. The $Cl(2,0)$ Clifford algebra admits $X$ and $Z$ as a gamma basis. An equivalent gamma basis can be expressed, e.g., through the ``entangled'' matrices ${\widetilde X}=\frac{1}{\sqrt{2}}(X+Z)$, ${\widetilde Z}=\frac{1}{\sqrt{2}}(X-Z)$.\par
In any case, due to the results in \cite{{oku},{crt}}, it is always possible to produce a $4$-character alphabetic
presentation of an irreducible gamma basis with words of given length $m$. In the Euclidean case $(q=0$), for instance, $m$ is explicitly given by the formula
\bea\label{4chlength}
m&=& \log_2G(k+1)+4r+1,
\eea
where $p\geq 2$ is parametrized according to
\bea
p&=& 8r+k+2,
\eea
with $r=1,2,\ldots$ and $k=0,1,2,3,4,5,6,7$, while $G(k+1)$ is given by the Radon-Hurwitz function
\cite{pt}
\bea
&
\begin{array}{|c|cccccccc|}\hline
n&1&2&3&4&5&6&7&8 \\ \hline
G(n)&1&2&4&4&8&8&8&8\\	\hline
\end{array}
\eea
The $mod~8$ property is in consequence of the famous Bott's periodicity.\par
We can therefore concentrate on the subclass of the alphabetic presentations, as previously defined.
Several questions can now be addressed. How many inequivalent alphabetic presentations can be defined?
The notion of the equivalence group should not be based of course on the class of similarity transformations connecting real-valued Clifford algebras, instead the notion of
a finite equivalence group of suitably defined moves transforming characters and words of an alphabetically
presented gamma basis into a new, equivalent, alphabetically presented gamma basis, should be given. Further questions can be addressed. Given the fact that $A$ is the only character whose square is negative ($A^2=-I$, with the (\ref{translation}) positions), any alphabetically presented Euclidean gamma basis (for $q=0$) admits words with even numbers of A's only (in the (\ref{alpha30}) case above $\gamma_3$ contains two A's, while $\gamma_1$, $\gamma_2$ contain no A's). Is it possible to define, for any $p$, Euclidean alphabetic presentations containing only the $3$ characters $I,X,Z$ (namely, limiting ourselves to a $3$-character alphabet)? Furthermore, which is the minimal length  ${\widetilde m}$ of the $3$-character Euclidean words? Under which conditions ${\widetilde m}$ coincides with $m$ given in (\ref{4chlength})? This is just a partial list of the questions that we are addressing (and partially solve) in this paper. To our knowledge, this type of program has never been investigated in the literature. Due to the recognized importance of Clifford algebras in several areas of mathematics and physics (for our purposes here it is sufficient to mention the applications to higher-dimensional unification theories like supergravities or superstrings \cite{gsw}, or the applications to robotics \cite{robotics}), we feel that it deserves being duly investigated. At the end of the paper we provide a very rough and preliminary list of possible topics which could benefit from it. The main core of the paper is devoted to the formulation of the problem and the presentation of general results and partial answers. The scheme of the paper is as follows. In the next Section we prove the Euclidean completeness of the $3$-character alphabetic presentations, introduce the equivalence group and a set of invariant functions. In Section $4$ we furnish a few algorithmic constructions to induce inequivalent $3$-character alphabetic presentations, compute the admissible invariants and present the results of an extensive computer search (for $3,4,5,6$-letter words). A table with the minimal lengths for $3$-character
alphabetic presentations of $Cl(p,0)$ is also given. Further issues and an outline of $4$-character alphabetic presentations will be discussed in the Conclusions. We will also mention there some topics which could benefit from the present investigation program.

\section{Alphabetic presentations}

In the Introduction we defined the alphabetic presentations of the gamma basis generators of a $Cl(p,q)$ Clifford algebra as given by $p+q$ words of $m$ letters constructed with the $4$ alphabetic characters
$I,A,X,Z$ (the alphabetic characters are in $1$-to-$1$ correspondence (\ref{translation}) with the four $2\times 2$ matrices (\ref{fundmat})). We also pointed out that, for Euclidean Clifford algebras ($q=0$) with $p\geq 2$, $3$-character
alphabetic presentations of the $p$ gamma basis generators could exist. Their words are constructed with the $I,X,Z$ characters alone. It is indeed easily proved that a $3$-character alphabetic presentation is Euclidean-complete. This means the following, for any $p$ it is always possible to find $p$ words satisfying (\ref{gammabasis}) and written with $I,X,Z$ alone. The completeness of the $4$-character alphabetic
presentations is guaranteed by the results given, e.g., in \cite{oku} and \cite{crt}. If the given $Cl(p,0)$ alphabetic
presentation contains no $A$'s, a $3$-character presentation immediately follows. If at least one word contains an $A$ in the $j$-th position, we can replace all $j$-th letter characters by two characters
(in $j$-th and $j+1$-th position) according to, for instance, $I\mapsto II$, $X\mapsto XX$, $Z\mapsto ZX$,
$A\mapsto IZ$.\footnote{This position leaves unchanged the anticommutation relations (\ref{gammabasis}) between two different characters. The square of $A$ changes sign. This, however, has no overall effect since each word of the Euclidean gamma basis contains an even number of $A$'s.} If the original words possess $m$ letters, the transformed words possess $m+1$ letters.
We can repeat the procedure every time we need to get rid of all $A$'s. The replacement leaves unchanged the (\ref{gammabasis}) relations. Applying the transformations to
the $Cl(3,0)$ gamma basis (\ref{alpha30}) we obtain, for instance, the $3$-character presentation
\bea\label{3ch30}
&\gamma_1 = IIXX,~~\gamma_2= IIZX,~~ \gamma_3 = IZIZ.&
\eea
It follows that a $3$-character presentation (not necessarily with minimal-length words) can always be found
for any $p$.
Translated back into the matrix language (tensor products of $2\times 2$ matrices), it produces representations of the (\ref{gammabasis}) $q=0$ generating relations in terms of matrices which are not necessarily irreducible. ``Alphabetic'' irreducibility should not be confused with matrix irreducibility.

\subsection{The alphabetic group of equivalence.}

We are now in the position to introduce the finite group of equivalence acting on alphabetic presentations. It is easier to discuss at first the $3$-character alphabetic presentations. It is
convenient to arrange the $p$ words of $m$ letters each of a given alphabetic $Cl(p,0)$ gamma basis
into a $p\times m$ rectangular matrix whose entries are the three alphabetic characters. The equivalence group $G$ acting on the $p\times m$ rectangular matrices is obtained by combining three types of moves:\\
{\em i}) permutations of the rows (they correspond to irrelevant reorderings of the $p$ words),\\
{\em ii}) permutations of the columns (the anticommutative property (\ref{gammabasis}) between two distinct
given words is unaffected by this operation),
\\
{\em iii}) transmutation of the characters in a given column: $X,Z$ are exchanged ($X\leftrightarrow Z$) while $I$ is unchanged (as before, the anticommutative property (\ref{gammabasis}) between two distinct
given words is unaffected by this operation).   \par
It should be noticed that the rectangular matrices can be simplified, without affecting the (\ref{gammabasis}) relations, by erasing the columns possessing entries with either a single character or the two characters
$I$ and $X$ or $I$ and $Z$ (the columns possessing both $X$ and $Z$ as entries cannot be erased). The process of erasing columns will be referred as ``simplification of the rectangular matrix''. A simple rectangular matrix is a rectangular matrix which cannot be further simplified. It produces a simple alphabetic presentation of a gamma basis. To be explicit, the (\ref{3ch30}) $3$-character presentation of the $Cl(3,0)$ gamma basis is associated to a $3\times 4$ rectangular matrix which can be simplified, erasing the first and the second columns, to produce a $3\times 2$ rectangular matrix according to
\bea\label{simple}
\begin{array}{cccc}
I&I&X&X\\
I&I&Z&X\\
I&Z&I&Z\\
\end{array}
	&\rightarrow&
\begin{array}{cc}
X&X\\
Z&X\\
I&Z\\
\end{array}
\eea
The simple rectangular matrix on the r.h.s. corresponds to three $2$-letter (length $2$) words. This is the minimal length for an alphabetic presentation of $Cl(3,0)$. It coincides with the minimal length of the (\ref{alpha30}) presentation which, on the other hand, requires $4$ characters instead of just $3$.\par
Two problems will be addressed in the next Section:\\
{\em 1}) which is the minimal length ${\widetilde m}$ of the words for a $3$-character alphabetic presentation of $Cl(p,0)$?\\
{\em 2}) how many inequivalent simple presentations of length $m$ can be found for a $3$-character alphabetic presentation of $Cl(p,0)$?\par
The second problem can be investigated with the help of invariants which detect the inequivalent classes under the finite group of transformations defined above. We introduce a few invariants, a ``horizontal invariant'' and the ``vertical invariants''.

\subsection{Alphabetic invariants.}

The horizontal invariant is defined as follows:
at first the number $m_I$ of $I$'s entries in any one of the $p$ rows is computed. Let us suppose we obtain
$i$ different results $k_1,\ldots, k_i$. We order them according to $k_1>k_2>\ldots >k_i\geq 0$.
Let $h_r$ be the number of rows producing the $k_r$ result ($r=1,2,\ldots, i$). Obviously
$h_1+h_2+\ldots+h_i=p$. The horizontal invariant $hor$ is expressed as an ordered set of the $h_r$ values with
$k_r$ as suffix. We write it as $hor({h_1}_{k_1}, {h_2}_{k_2}, \ldots, {h_i}_{k_i})$. It is easily checked that $hor$ is invariant under the group transformations (permutations and transmutations).
As an example, the $hor$ invariant of the simple rectangular matrix in the r.h.s. of (\ref{simple})
is $hor(1_1,2_0)$.\par
The first vertical invariant $ver$ is analogously defined; the difference is that the number $n_I$ of $I$'s entries is computed in terms of the columns. Let us suppose we get $j$ different results $l_1, \ldots, l_j$, ordered according to $l_1>l_2>\ldots > l_j\geq 0$. Let $v_r$ be the number of columns producing the $l_r$ result ($v_1+v_2+\ldots+v_j=m$). The vertical invariant $ver$ is expressed as
$ver({v_1}_{l_1},{v_2}_{l_2},\ldots,{v_j}_{l_j})$. The $ver$ invariant of the simple rectangular matrix in the r.h.s. of
(\ref{simple}) is explicitly given by $ver(1_1,1_0)$.\par
The second vertical invariant $\widetilde{ver}$ is defined as $ver$, but instead of counting the number $n_I$ of $I$'s in a given column, we compute the absolute difference $n_{XZ}=|n_X-n_Z|$ between the number of $X$'s and the number of $Z$'s entries in any given column. Applied to (\ref{simple}) we obtain ${\widetilde{ver}}(1_1,1_0)$. A less refined invariant under the group generated by permutations and transmutations is the total number $N_I$ of $I$'s entries in a simple rectangular matrix.\\
A more refined invariant is ${\widehat{ver}}$, counting the number $v_{(k_i,l_i)}$ of columns presenting the given pair $(n_I=k_i,n_{XZ}=l_i)$. The result is presented as
${\widehat{ver}}(v_{(k_1,l_1)},v_{(k_2,l_2)},\ldots , v_{(k_j,l_j)})$ (the pairs are conveniently ordered). Applied to (\ref{simple}) we obtain ${\widehat{ver}}(1_{(1,0)},1_{(0,1)})$.\\
In the next Section it is sufficient to use the invariants $hor$ and $ver$ (based on the counting of $I$'s) to detect the inequivalent $3$-letter and $4$-letter alphabetic presentations.\\ 
For $m=1$ (single-letter words) we have a unique $Cl(2,0)$ gamma basis given by $\{X,Z\}$.\\
For $m=2$ we have four equivalent (under permutations and transmutations) presentations of
$Cl(3,0)$, given by
$\{XX, ZX,IZ\}$, $\{XX, XZ, ZI\}$, $\{ZZ,XZ, IX\}$, $\{ZZ, ZX, XI\}$.\par
In the next Section we discuss the construction of $3$-character alphabetic presentations with $m$-letter words for higher values of $m$.

\section{Inequivalent $3$-character alphabetic presentations}

In the previous Section we furnished the $m$-letter $3$-character alphabetic presentations for $m=1,2$.
We discuss now the situation for $m\geq 3$. In order to do that, besides the already introduced notion of ``simple alphabetic presentation'', we also need to define the notion of ``maximally extended alphabetic
presentation''. It corresponds to an $m$-letter gamma basis $B$ such that no further word, anticommuting with all the words in $B$, can be added (in the following, explicit examples of non-maximally extended gamma basis will
be given; they are obtained by erasing at least one word from a maximally extended gamma basis). It turns out that, at any given $m$, the classification of the inequivalent gamma basis is recovered from the classification
of the simple, maximally extended, gamma basis.\par
In \cite{crt} an algorithmic presentation was given to induce new gamma basis from previously known ones.
In a very simple form (which is applied to the Euclidean case), it corresponds to produce an $(m+1)$-letter
gamma basis for the $Cl(p+1,0)$ Clifford algebra in terms of an $m$-letter gamma basis for $Cl(p,0)$. If we denote with $\gamma_i$ the words in the $Cl(p,0)$ gamma basis, it is sufficient to express the ${\widetilde
\gamma}_j$ words ($j=1,2,\ldots,p+1)$ in the $Cl(p+1,0)$ gamma basis as
\bea\label{algo0}
{\widetilde \gamma}_i&=& \gamma_iX,\nonumber\\
{\widetilde\gamma}_{p+1}&=& I^{(m)}Z,\quad\quad I^{(m)}\equiv II\ldots I ~(taken~m~times).
\eea
It is easily shown that the above position in general does not exhaust the class of inequivalent (in the sense specified in the previous Section) $(m+1)$-letter alphabetic presentations of $Cl(p+1,0)$. A general algorithm can be presented through the following construction. Let ${\cal A},{\cal B}_1,{\cal B}_2$ be three sets of $m$-letter
words (whose respective cardinalities are $n_A,n_{B_1},n_{B_2}$) satisfying the following properties: both ${\cal C}_1={\cal A}\cup{\cal B}_1 $ and ${\cal C}_2={\cal A}\cup{\cal  B}_2$ are a gamma basis and,
furthermore, the words in ${\cal B}_1$ commute with all the words in ${\cal B}_2$. Under these conditions an $(m+1)$-letter
presentation ${\widetilde{\cal B}}$ of a $Cl(n_A+n_{B_1}+n_{B_2},0)$ gamma basis can be produced by setting, symbolically,
\bea\label{algo1}
{\widetilde {\cal B}}&\equiv& \{{\cal A}I,{\cal B}_1X,{\cal B}_2Z\}.
\eea
One should notice that ${\cal A}$ could be the empty set while both ${\cal B}_1,{\cal B}_2$ must necessarily be non-empty in order for ${\widetilde{\cal B}}$ to be a simple gamma basis.\par
We applied this algorithm to induce, for $m=3,4$, the whole set of inequivalent, simple, maximally extended, gamma basis. In parallel we produce a systematic computer search of the inequivalent, simple, maximally extended gamma basis for $m=3,4,5,6$. The results are reported below.

\subsection{$3$-letter alphabetic presentations.}

For $m=3$ there are only three inequivalent, simple, maximally extended gamma basis (two for $p=4$, one for $p=5$). The representatives in each given class and their associated invariants are explicitly given by
\bea\label{m3}
&\begin{array}{ccl}
{\underline{4_\alpha}}~(p=4):&
\begin{array}{ccc}
X&X&X\\
X&X&Z\\
X&Z&I\\
Z&I&I\\
\end{array}
& [~ hor(1_2,1_1,2_0);~ ver(1_2,1_1,1_0);~ N_I=3~],\\
&&\\
{\underline{4_\beta}}~(p=4):&
\begin{array}{ccc}
X&X&I\\
X&Z&I\\
Z&I&X\\
Z&I&Z\\
\end{array}
& [ ~hor(4_1);~ ver(2_2,,1_0);~ N_I=4~],\\
&&\\
{\underline 5}~(p=5):&
\begin{array}{ccc}
X&X&X\\
X&I&Z\\
I&Z&X\\
Z&X&I\\
Z&Z&Z
\end{array}
& [ ~hor(3_1,2_0);~ ver(3_1);~ N_I=3~].\end{array}&
\eea
One should notice that two inequivalent $p=4$ non-maximally extended gamma basis are obtained by erasing one word from ${\underline 5}$; if the word to be erased is $XXX$, we obtain a gamma basis with horizontal invariant $hor(3_1,1_0)$
while, if the word to be erased is $XIZ$, we obtain a gamma basis
($\{XXX,IZX,ZXI,ZZZ\}$)
with horizontal invariant $hor(2_1,2_0)$.
The $N_I$ invariant of the first case (the $\{XIZ,IZX,ZXI,ZZZ\}$ gamma basis) is $N_I=3$, which means that it
is not sufficiently refined to detect a difference between this non-maximally extended representation and the maximally extended ${\underline{4_\alpha}}$ gamma basis. Erasing from both cases above an extra, conveniently chosen,
word, we produce two inequivalent $p=3$ simple non-maximally extended gamma basis. They are given by
$\{XIZ, ZXI,IZX\}$, with horizontal invariant $hor(3_1)$, and $\{XXX,XIZ,ZZZ\}$ with $hor(1_1,2_0)$.\par
On the other hand, erasing a word from either ${\underline{4_\alpha}}$ or ${\underline{4_\beta}}$, produces in both cases a $p=3$ non simple gamma basis.
\par
It is quite illustrative to show how ${\underline{4_\alpha}},{\underline{ 4_\beta}},{\underline{ 5}}$ in  (\ref{m3}) can be algorithmically computed in terms of (\ref{algo1}). We get
\bea
\left(\begin{array}{lll}
{\cal A}&=&\{IZ\}\\
{\cal B}_1&=&\{ZX,XX\}\\{\cal B}_2&=&\{IX\} \end{array}\right)&\Rightarrow& \{IZI,ZXX,XXX,IXZ\}\in {\underline{4_\alpha}},\nonumber\\
\left(\begin{array}{lll}
{\cal A}&=&\{XX,XZ\}\\
{\cal B}_1&=&{\cal B}_2~=\{IZ\} \end{array}\right)&\Rightarrow& \{XXI,ZXI,IZX,IZZ\}\in {\underline{4_\beta}},\nonumber\\
\left(\begin{array}{lll}
{\cal A}&=&\{XX\}\\
{\cal B}_1&=&\{ZX,IZ\}\\ {\cal B}_2&=&\{XZ,ZI\} \end{array}\right)&\Rightarrow& \{XXI,ZXX,IZX, XZZ,ZIZ\}\in {\underline{5}}.
\eea

\subsection{$4$-letter alphabetic presentations.}

Starting from $m\geq 4$ a new feature arises. Simple, maximally extended gamma basis with non-minimal length words are produced. Indeed, four inequivalent such representations for $p=5$ can be found. On the other hand, as we have seen, a $p=5$ gamma basis is already encountered for $m=3$. Translated back into matrix representations, the four $p=5$, $m=4$ gamma basis produce reducible (in matrix, not alphabetic, sense) $16\times 16$ gamma matrices whose size is twice the $8\times 8$ irreducible representation obtained from ${\underline 5}$ in (\ref{m3}). The representatives of the four inequivalent $p=5$, $m=4$ gamma basis and their associated invariants are explicitly given by
\bea\label{m4red}
&\begin{array}{ccl}
{\underline{5_\alpha}}~(p=5):&
\begin{array}{cccc}
X&X&X&X\\
X&X&I&Z\\
I&Z&X&I\\
Z&I&Z&Z\\
Z&Z&Z&X
\end{array}
& [ ~hor(1_2,2_1,2_0);~ ver(4_1);~ N_I=4~],\\
&&\\
{\underline{5_\beta}}~(p=5):&
\begin{array}{cccc}
X&X&X&X\\
X&X&X&Z\\
X&X&Z&I\\
X&Z&I&I\\
Z&I&I&I
\end{array}
& [~ hor(1_3,1_2,1_1,2_0);~ ver(1_3,1_2,1_1,1_0);~ N_I=6~],\\
&&\\
{\underline{5_\gamma}}~(p=5):&
\begin{array}{cccc}
X&X&X&I\\
X&X&Z&I\\
X&Z&I&X\\
X&Z&I&Z\\
Z&I&I&I
\end{array}
& [~ hor(1_3,4_1);~ ver(2_3,1_1,1_0);~ N_I=7~],\\
&&\\
{\underline{5_\delta}}~(p=5):&
\begin{array}{cccc}
X&X&X&I\\
X&X&Z&I\\
X&Z&I&I\\
Z&I&I&X\\
Z&I&I&Z
\end{array}
& [~ hor(3_2,2_1);~ ver(2_3,1_2,1_0);~ N_I=8~].\end{array}&
\eea
$m=4$ is the minimal length for an alphabetic presentation of the Euclidean Clifford algebra with $p=6,7,8$.
The complete list (representatives and their associated invariants) of inequivalent, simple, maximally extended gamma basis for $p=6,7,8$ and $m=4$ is explicitly given by
\bea\label{m4irred}
&\begin{array}{ccl}
{\underline{6_\alpha}}~(p=6):&
\begin{array}{cccc}
X&X&X&X\\
Z&I&X&X\\
X&Z&I&X\\
I&X&Z&X\\
Z&Z&Z&X\\
I&I&I&Z
\end{array}
& [~ hor(1_3,3_1,2_0);~ ver(3_2,1_0);~ N_I=6~],\\
&&\\
{\underline{6_\beta}}~(p=6):&
\begin{array}{cccc}
X&X&X&X\\
Z&X&X&X\\
I&Z&I&X\\
I&X&Z&I\\
I&I&X&Z\\
I&Z&Z&Z
\end{array}
& [ ~hor(3_2,1_1,2_0);~ ver(1_4,3_1);~ N_I=7~],\\
&&\\
{\underline{6_\gamma}}~(p=6):&
\begin{array}{cccc}
I&X&X&X\\
I&Z&X&X\\
X&I&Z&X\\
Z&I&Z&I\\
Z&I&X&Z\\
X&I&I&Z
\end{array}
& [ ~hor(2_2,4_1);~ ver(1_4,1_2,2_1);~ N_I=8~];\\
&&\\
{\underline 7}~(p=7):&
\begin{array}{cccc}
X&X&X&X\\
Z&I&X&X\\
X&Z&I&X\\
I&X&Z&I\\
Z&Z&Z&I\\
I&I&X&Z\\
X&Z&Z&Z
\end{array}
& [ ~hor(2_2,3_1,2_0);~ ver(3_2,1_1);~N_I=7~];\\
&&\\
{\underline 8}~(p=8):&
\begin{array}{cccc}
I&X&X&X\\
X&Z&I&X\\
Z&I&Z&X\\
Z&Z&X&I\\
X&X&Z&I\\
X&I&X&Z\\
Z&X&I&Z\\
I&Z&Z&Z
\end{array}
& [~ hor(8_1);~ ver(4_2);~ N_I=8~].\end{array}&
\eea
All the gamma basis entering (\ref{m4red}) and (\ref{m4irred}) can be algorithmically produced
with the (\ref{algo1}) construction. For simplicity we limit ourselves
to present the algorithmic construction of the largest of such representations, the gamma basis ${\underline 8}$ in (\ref{m4irred}) which
generates $Cl(8,0)$. The sets ${\cal A},{\cal B}_1,{\cal B}_2$ are given by
\bea\label{n1cyclic}
&& {\cal A}=\{XXX,ZZZ\},\quad
{\cal B}_1=\{ZXI,XIZ,IZX\}\quad {\cal B}_2=\{XZI,ZIX,IXZ\}\quad \Rightarrow \nonumber\\&&
\Rightarrow \{XXXI,ZZZI,
ZXIX,XIZX,IZXX,XZIZ,ZIXZ,IXZZ\}\in {\underline 8}.
\eea
This is the first example of the subclass of ``cyclic'' algorithmic constructions that will be discussed later.\par
We made an exhaustive computer search and listed all inequivalent, simple, maximally extended, $3$-character alphabetic presentations for $m=5$ and $m=6$. To save space we just limit ourselves to mention that $5$-letter
words can produce an Euclidean gamma basis for at most $p=9$ while $6$-letter words can produce a gamma basis for at most $p=10$.

\subsection{The minimal lengths.}

We are now in the position to present a table with the minimal length ${\widetilde m}$ required to produce a
$3$-character alphabetic presentation of $Cl(p,0)$ at a given $p$. We compare ${\widetilde m}$ with ${m}$,
the minimal length for $4$-character alphabetic presentations, given by (\ref{4chlength}). We get
\bea\label{3lettml}
&
\begin{array}{|c|c|c|cc|cccc|c|c|cc|cccc|c|c|}\hline
p&2&3&4&5&6&7&8&9&10&11&12&13&14&15&16&17&18&\ldots\\ \hline
m&1&2&3&3&4&4&4&4&5&6&7&7&8&8&8&8&9&\ldots\\	\hline
{\widetilde m}&1&2&3&3&4&4&4&5&6&7&7&8?&8&8&8&9?&10?&\ldots\\	\hline
\end{array}&
\eea

The ${\widetilde m}$ values for $p=13,17,18$ are conjectured since a formal proof is lacking.\par
The above table is the result of an explicit computer search for ${\widetilde m}\leq 6$, combined with
algorithmic constructions for ${\widetilde m}>6$.

\subsection{The cyclic prescription and another algorithm.}

There is a class of gamma basis (let us call them ``cyclic''), obtained by a specific choice of ${\cal A},{\cal B}_1,{\cal B}_2$ entering (\ref{algo1}).\par
For integral values $n=1,2,3,\ldots$, we construct ${\cal B}_1$ as a set of $2n+1$ words of $(2n+1)$-length obtained by cyclically
permuting $IZXZX\ldots ZX\equiv I(ZX)^{(n)}$, while ${\cal B}_2$ is the set of $2n+1$ words of $(2n+1)$-length obtained by cyclically
permuting $IXZXZ\ldots XZ\equiv I(XZ)^{(n)}$:
\bea\label{b1b2}
{\cal B}_1&=&\{I(XZ)^{(n)} ~and~its ~cylic~permutations\},\nonumber\\
{\cal B}_2&=&\{I(ZX)^{(n)}~and~its~cyclic~permutations\}.
\eea
Clearly, the words in ${\cal B}_1$ commute with the words in ${\cal B}_2$.\par
Two subcases are now considered:\\
{\em subcase~i}) for odd values $n$, ${\cal A}$ is given by the $2$ words set
\bea\label{aodd}
{\cal A}&=&\{Z(ZZ)^{(n)}, X(XX)^{(n)}\};
\eea
{\em subcase~ii}) for even values $n$, $A$ is the empty set
\bea\label{aeven}
{\cal A}&=&\oslash.
\eea \par
The (\ref{algo1}) prescription gives us, in both cases, a $(2n+2)$-letter gamma basis such that:\\
{\em subcase~i}) for odd values $n$, $p= 2\cdot(2n+1)+2=4(n+1)$ and\\
{\em subcase~ii})
for even values $n$, $p=2\cdot(2n+1)=4n+2$.\par
For $n=1$ we recover the construction of the ${\underline 8}$ gamma basis given in (\ref{n1cyclic}).\par
 As a result we obtain a relation, for cyclic $3$-character representations, between $p$ and the length
${\overline m}$ of their words, given by
\bea
&
\begin{array}{|c|c|c|c|c|c|}\hline
p&8&10&16&18&\ldots\\ \hline
{\overline m}&4&6&8&10&\ldots\\	\hline
\end{array}&
\eea
We know that in the subcase {\em i}, for $p=8k$ ($k=1,2,\ldots $), ${\overline m}=4k$ is a minimal length because it coincides with the known minimal length for $4$-character presentations. On the other hand we explicitly checked  that in the subcase {\em ii}, for $p=10$, ${\overline m}=6$ corresponds to a minimal length while the subcase {\em ii} provides an upper bound for the minimal length for $p=18$.\par
Another algorithmic construction, different from the cyclic prescription and generalizing the (\ref{algo0}) algorithm, allows us to prove that ${\widetilde m}=7$ in (\ref{3lettml}) is indeed the $3$-character
minimal length for $p=12$. Let $C_1, C_2$ be two gamma basis for, respectively, $Cl(p_1,0), Cl(p_2,0)$ with $m_1,m_2$ length of their words. Let $\gamma$ be a word of $C_1$ and ${\widehat{C_1}}$ the complement set of
$\{\gamma\}$ in $C_1$. A new gamma basis $C$ for $Cl(p_1+p_2-1,0)$, with words of length $m=m_1+m_2$, is symbolically given by
\bea\label{algo2}
C&=& \{ \gamma C_2 , {\widehat{C_1}}I^{(m_2)}\}.
\eea
By taking, e.g., ${\underline 5}$ in (\ref{m3}) as $C_1$ and ${\underline 8}$ in (\ref{m4irred}) as $C_2$ we obtain a $3$-character gamma basis with $p=5+8-1=12$ and $m=3+4=7$.

\section{Conclusions and outlook}

In this work we investigated the alphabetic representations of the $Cl(p,q)$ Clifford algebras gamma basis. The gamma basis generators are expressed as words written in a up to four characters alphabet. The four characters, $I,X,Z,A$, are associated with four $2\times 2$ matrices according to (\ref{translation}) ($I$ corresponds to the identity matrix, $A$ to the antisymmetric matrix, etc.) and satisfy the anticommutation relations (\ref{gammabasis}). The words of an alphabetic representation are in correspondence with the matrix tensor products (in the correspondence, the tensor product symbol is omitted).  \par
The interesting alphabets to consider are the whole $4$-character alphabet or a $3$-character alphabet. A $2$-character alphabet given by, e.g., $X$ and $Z$, is too poor; indeed, it can only produce an Euclidean gamma basis for $p=1,2$. On the other hand, the $3$-character alphabet given by $I,X,Z$ is Euclidean-complete. It produces $Cl(p,0)$ Euclidean gamma basis for any value of $p$. For this alphabet we introduced the notion of the alphabetic group of equivalence, constructed invariants and derived general and
partial results (concerning, e.g., the minimal length of the words which produce a gamma basis for a given
$p$). The alphabetic group of equivalence $G$ can be extended to the whole $4$-character alphabet or to a $3$-character alphabet containing $A$ (namely, the character associated with the antisymmetric matrix). $G$
is based on three types of moves, the permutations (of rows and columns) and the transmutations of characters.
In the extended case the transmutations have to be suitably restricted, since an $A\leftrightarrow X$ (or an $A\leftrightarrow Z$) transmutation maps a $Cl(p,q)$ gamma basis into a $Cl(p',q')$ gamma basis (the constraint $p'+q'=p+q$ is satisfied; in the general case, on the other hand, $p'$ differs from $p$). A viable restriction in the definition of the alphabetic group of equivalence consists in disregarding the transmutations involving the $A$ character. Besides the invariants discussed in the subsection $2.2$, extra horizontal and vertical invariants, counting the number of the $A$'s character, have to be introduced. The analysis of the $4$-character case (invariants, inequivalent alphabetic presentations, etc.) is left for forthcoming publications. It is worth pointing out that the introduction of a
fourth character greatly increases the time needed for computer search of the inequivalent alphabetic
presentations.\par
To our knowledge this investigation program has not been addressed in the literature. We have proven here that it is based on a well-posed mathematical problem admitting interesting and quite non-trivial solutions.\par We have postponed so far discussing its possible applications. In the light of this we should mention that the whole idea of constructing and analyzing the alphabetic presentations was deeply
rooted in the investigations in our respective fields. Clifford algebras (in their alphabetic presentations) are the basis to construct \cite{krt} representations of the $N$-extended supersymmetric quantum mechanics. These representations are nicely encoded in a graphical interpretation (see \cite{fg} and \cite{kt}) in terms of colored, oriented, graphs. The equivalence group of transformations acting on graphs is related with the alphabetic group of transformations of the associated Clifford algebra.\par
The applications of Clifford algebras to robotics have been detailed, e.g., in \cite{robotics}. An interesting possibility is offered by the construction of cellular automata which manipulate words in an alphabetic presentation of Clifford algebras.
The $3$-character alphabet, here investigated in detail, is the simplest of such settings which allows the necessary complexity
(Euclidean completeness, inequivalent alphabetic representations, etc.).\par At the end let us just mention a seemingly far-fetched possibility which, nevertheless, we believe deserves being duly investigated. The DNA codon problem concerns the yet to be explained degeneracies found in associating aminoacids with the triplets of the DNA nucleotides, cytosine (C), adenine (A), thymine (T), guanine (G) for DNA or their respective
G, U (for uracil), A, C complements for mRNA. In the vertebral mitochondrial code, for instance, the $4^3=64$ nucleotides triples are associated to 20 aminoacids and a stop signal according to a decomposition assigning $2$, $4$ or $6$ different words to each aminoacid and the stop signal: $64= 2\times 6+7\times 4+12\times 2$. One can consult \cite{dd} for an updated discussion of the codon problem and
the attempted solutions (based on p-adic distance, deformed superalgebras, etc.). It is quite tempting to reformulate this problem in terms of alphabetic presentations of Clifford algebras (identifying each nucleotide with one of the four characters $I$, $X$, $Z$ and $A$) and check whether the alphabetic invariants could play a role in the association with the aminoacids.
\par
{}~
\\{}~{}
\par {\large{\bf Acknowledgments}}{} ~\\{}~\par
F.T. is grateful to Rold\~ao da Rocha for useful comments. \par
This work has been supported by Edital Universal CNPq, Proc. 472903/2008-0.

\end{document}